\documentclass[prl, reprint]{revtex4-1}

\usepackage{amsmath}    % need for subequations
\usepackage{graphicx}   % need for figures
\usepackage{verbatim}   % useful for program listings
\usepackage{color}      % use if color is used in text
\usepackage{subfigure}  % use for side-by-side figures
\usepackage{hyperref}   % use for hypertext links, including those to external documents and URLs
\usepackage{amssymb}
\usepackage{caption}

\usepackage{epstopdf}

\begin{document}

\title{Topological States with Broken Translational and Time-Reversal Symmetries in a Honeycomb-Triangular Lattice}
\author{R.\ Sarjonen and P.\ T\"{o}rm\"{a}}
\affiliation{COMP Centre of Excellence, Department of Applied Physics, Aalto University, FI-00076 Aalto, Finland}
\date{\today}

\begin{abstract}
We study fermions in a lattice, with on-site and nearest neighbor attractive interactions between two spin species. We consider two geometries: both spins in a triangular lattice, and a mixed geometry with up-spins in honeycomb and down-spins in triangular lattices. We focus on the interplay between spin-population imbalance, on-site and valence bond pairing, and order parameter symmetry. The mixed geometry leads to a rich phase diagram of topologically non-trivial phases. In both geometries, we predict order parameters with simultaneous time-reversal and translational symmetry breaking.
\end{abstract}

\maketitle

In the Bardeen, Cooper and Schrieffer (BCS) theory \cite{Bardeen}, which describes well many low temperature superconductors, the transition to the superconducting state is characterized by the breaking of gauge symmetry only. However, the hallmark of unconventional superconductivity is the breaking of additional symmetries. For example, the Fulde-Ferrel-Larkin-Ovchinnikov (FFLO) state has broken translational symmetry: the order parameter has a non-trivial spatial dependence \cite{Fulde1,Larkin,Jin,Roy}. On the other hand, chiral superconductors break time-reversal symmetry (TRS) because they feature gap parameters that wind in phase around the Fermi surface in multiples of $2\pi$. Chiral superconductors also exhibit many other fascinating properties that are highly sought after for nanoscience applications \cite{Nandkishore, chiralSC1, chiralSC2, chiralSC3, chiralSC4}, and broken TRS is a prerequisite for the quantum Hall effects (excluding the spin Hall effect) \cite{Kane,Bernevig}. Moreover, in MgB$_2$ and iron pnictides \cite{ironSC0, ironSC1,ironSC2,ironSC3} TRS may be broken due to interband couplings \cite{Huang1,Hu,Lin1}. In this letter, we propose and theoretically study a system in which exotic superfluids with translational and TRS breaking can compete and even coexist.

Simultaneous breaking of multiple symmetries is an intriguing phenomenon; an example of a fervently sought after state is the supersolid which breaks translational and $U(1)$ symmetries by coexisting crystal structure and superfluidity \cite{Boninsegni}.
%[REFERENSSI RevModPhys 84, 759 (2012)] 
As another example, it was recently predicted for spinless fermions in a triangular lattice that density orders with several broken symmetries may coexist \cite{Eckardt}. Each broken symmetry typically generates characteristic modes, the coexistence of which leads to rich physics and potentially applications. Achieving such states is, however, non-trivial since the system must be susceptible to different types of order. The novel translational and TRS breaking superfluids that we predict here are of conceptual interest as a new type of state with simultaneous breaking of several symmetries, all reflected in the superfluid order parameter. Importantly, the very ingredients that are essential for creating such states, namely a combination of long-range interactions, special lattice geometries and spin-density imbalance, are an emerging experimental reality in ultracold gas systems.   
  
A crucial extension to the capabilities of ultracold Fermi gases as a quantum simulator \cite{Houches,FHPhys}, including emulation of the extended Fermi-Hubbard model \cite{LiangHe,Gorshkov,Bhongale,Kiesel,Huang2}, is emerging from the new possibilities of realizing not only on-site but also long-range interactions. They can be realized, for example, with the help of atoms with a large magnetic dipole moment (e.g., chromium, dysprosium and erbium \cite{Pfau, Lev, Ferlaino}), dipolar molecules such as the fermionic $^{40}$K$^{87}$Rb \cite{Ye, NatureBo,NatureNi}, or atoms excited to Rydberg states \cite{Molmer,Daley,Raithel,Bloch}. Another type of possibility are mixtures of bosonic and fermionic atoms where the bosons induce a long-range interaction between the fermions \cite{Lim2}.
%[TAHAN SUORA REFERENSSI VIITTEESEEN 57] 
Intriguingly, ultracold gas lattice systems also enable spin-dependent confinement of particles \cite{Soltan-Panahi,WindPassinger,Thywissen,Minardi}. This has led to theoretical proposals of new concepts, such as mixed geometry pairing \cite{Topol}.  

\begin{figure}[h!]
\centering
\subfigure{\includegraphics[scale=0.32]{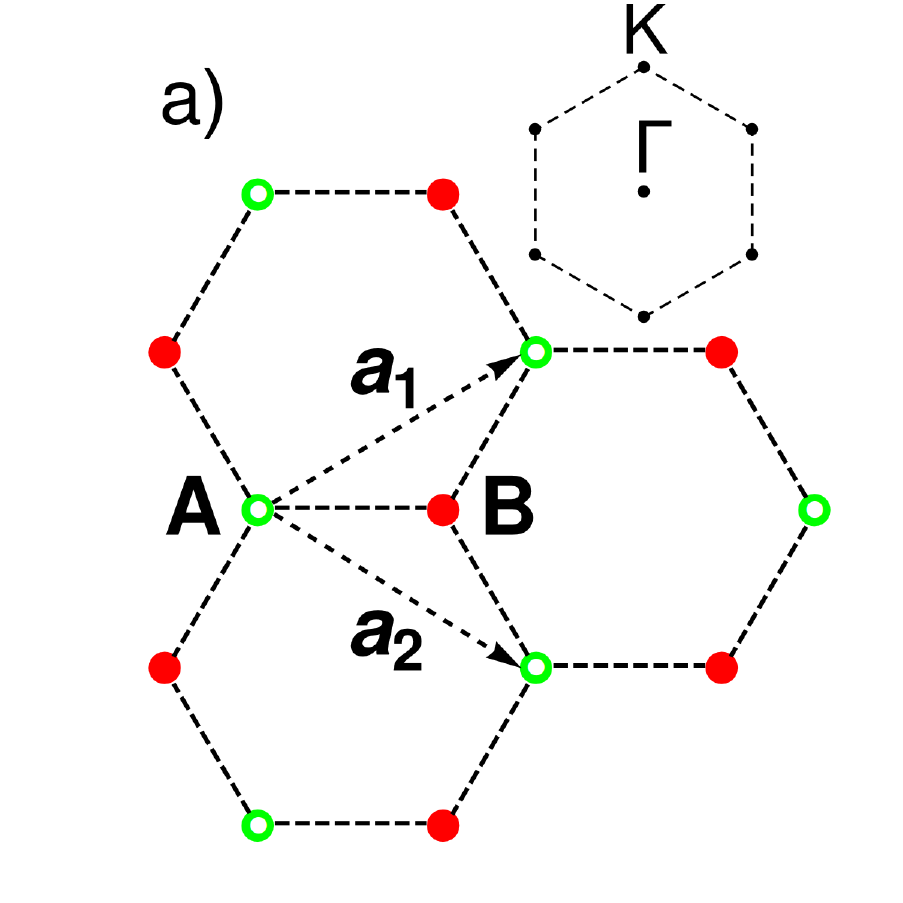} \label{fig1a}}
\subfigure{\includegraphics[scale=0.245]{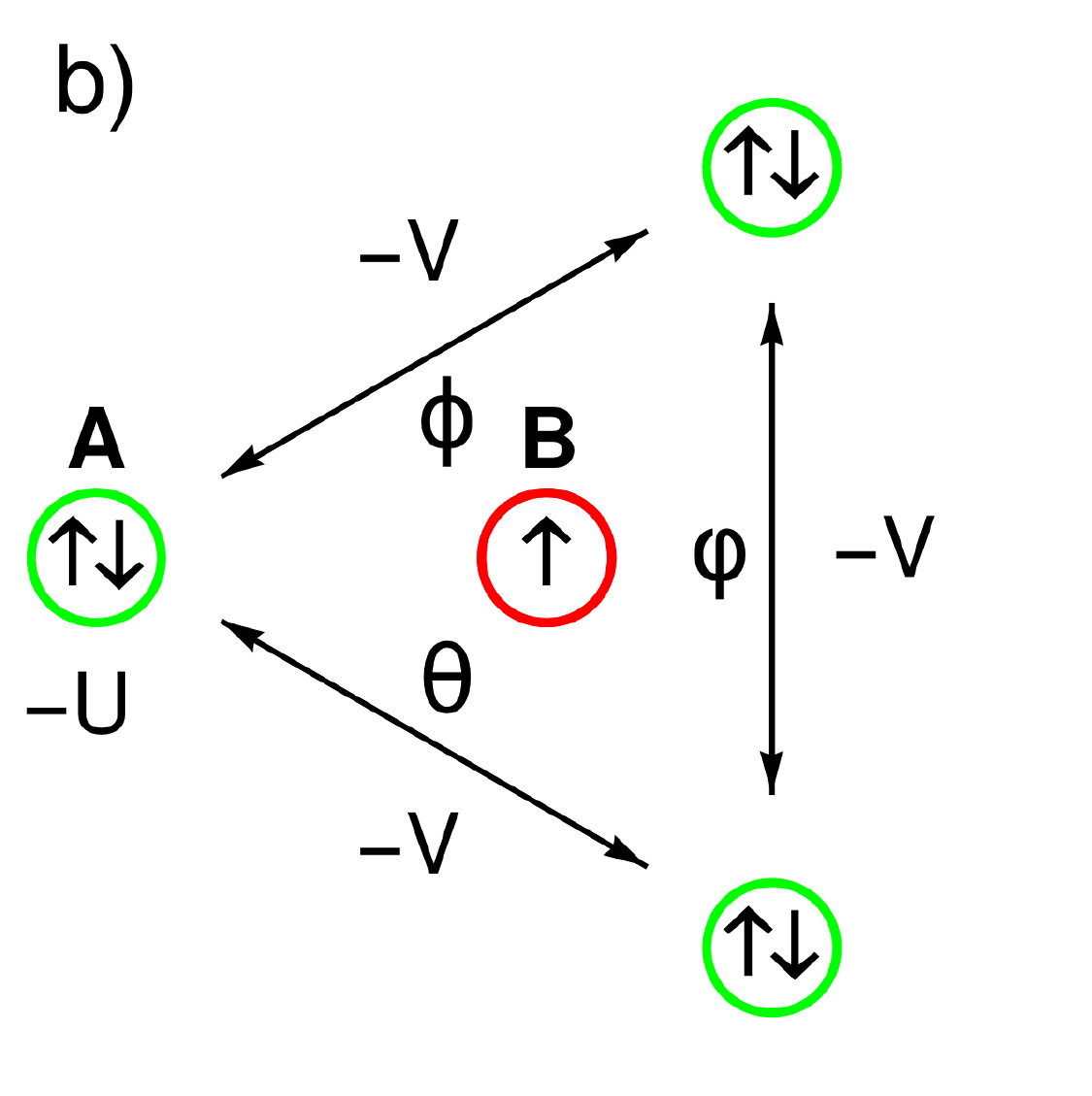} \label{fig1b}}
\caption{\label{fig1}(Color online.) a) Honeycomb-triangular lattice and its Brillouin zone. Sublattice A (green) houses both spins, whereas sublattice B (red) only up-spin particles. b) Pairing happens both on-site and inter-site with energy gains $-U$ and $-V$, respectively. Phases of the bond order parameters are marked with $\theta$, $\phi$ and $\varphi$.}
\end{figure}

We consider two different lattice systems, namely a honeycomb-triangular and a triangular lattice loaded with spin-1/2 fermions. In the former system, the honeycomb lattice comprises two triangular sublattices A and B as shown in Fig.\ \ref{fig1a}. The sublattices are spin-selective in such a way that $\uparrow$-spin atoms can occupy the whole honeycomb lattice, but $\downarrow$-spin atoms are confined to the triangular sublattice A. Consequently, we denote the honeycomb lattice by $\mathcal{L_{\uparrow}}$ and the triangular sublattice A by $\mathcal{L_{\downarrow}}$.

We assume that $\uparrow$-spin and $\downarrow$-spin atoms can tunnel only between neighboring sites of $\mathcal{L_{\uparrow}}$ and $\mathcal{L_{\downarrow}}$, respectively. We denote the tunneling amplitudes of $\uparrow$-spin and $\downarrow$-spin atoms by $t_{\uparrow}$ and $t_{\downarrow}$, respectively. Subsequently, the Hamiltonian that takes into account tunneling and possible on-site energy modulations can be written as
\begin{eqnarray}
\hspace{-3mm}\mathcal{H}_0&=&-t_{\uparrow} \sum_{\langle i,j\rangle\in\mathcal{L}_{\uparrow}} 
( \hat{a}_{i\uparrow}^{\dagger}\hat{b}_{j\uparrow}^{\vphantom{\dagger}} +\mbox{H.c.} )
-\mu_{\uparrow} \sum_i (\hat{n}_{i\uparrow}^a+\hat{n}_{i\uparrow}^b )
\nonumber \\
&&{}-t_{\downarrow} \sum_{\langle i,j\rangle\in\mathcal{L}_{\downarrow}}( \hat{a}_{i\downarrow}^{\dagger}\hat{a}_{j\downarrow}^{\vphantom{\dagger}} +\mbox{H.c.} )
-(\mu_{\downarrow} - \epsilon_{\downarrow}^a) \sum_i \hat{n}_{i\downarrow}^a, 
\end{eqnarray}
where $\hat{a}^{\dagger}$ ($\hat{a}$) and $\hat{b}^{\dagger}$ ($\hat{b}$) are fermionic creation (annihilation) operators in sublattices A and B, respectively, and $\hat{n}^a$ and $\hat{n}^b$ are the corresponding density operators. Parameters $\mu_{\uparrow}$ and $\mu_{\downarrow}$ are chemical potentials for $\uparrow$-spin and $\downarrow$-spin particles, respectively. We choose $\epsilon_{\downarrow}^a=-3$, and set $t_{\uparrow}=t_{\downarrow}=t=1$ in all our calculations. For the triangular lattice the Hamiltonian is otherwise the same, but there are no B site energy modulation terms and $\uparrow$-spin tunneling happens between neighboring A sites (see Supplemental Material \cite{Omasupp}).

We also consider attractive on-site and nearest-neighbor (NN) interactions. The on-site interaction takes place at A sites, and we denote the interaction strength by $-U$ where $U\geq0$. Subsequently, the corresponding Hamiltonian reads
\begin{equation}
\mathcal{H}_{\mbox{\scriptsize{os}}} = -U\sum_j \hat{n}_{j\uparrow}^a \hat{n}_{j\downarrow}^a.
\end{equation}
In conventional superconductivity, electrons form superconducting Cooper pairs in a spin-singlet state \cite{Fossheim}. However, spin-singlet bonding between neighboring A and B sites is impossible because $\downarrow$-spin particles cannot occupy B-sites. Therefore we assume that the nearest-neighbor interaction takes place between adjacent A sites and represent it with the Hamiltonian (see Supplemental Material Section I.E.2)
\begin{equation}
\mathcal{H}_{\mbox{\scriptsize{nn}}} = -V\sum_{\langle m,n \rangle\in\mathcal{L}_{\downarrow}}
\hat{h}_{mn}^{\dagger} \hat{h}_{mn},
\end{equation}
where $\hat{h}_{mn}^{\dagger}=(\hat{a}_{m\uparrow}^{\dagger} \hat{a}_{n\downarrow}^{\dagger} - \hat{a}_{m\downarrow}^{\dagger} \hat{a}_{n\uparrow}^{\dagger})/\sqrt{2}$ is a spin-singlet creation operator. The parameter $V>0$ represents an energy gain when two atoms form a spin-singlet bond, because $\hat{h}_{mn}^{\dagger} \hat{h}_{mn}$ is the number operator for singlet bonds \cite{Baskaran}. We note that the spin-singlet states between neighboring sites are essentially resonating-valence-bond (RVB) states proposed by Anderson \cite{Anderson}.

The full Hamiltonian is
\begin{equation}
\mathcal{H}=\mathcal{H}_0 + \mathcal{H}_{\mbox{\scriptsize{os}}} + \mathcal{H}_{\mbox{\scriptsize{nn}}}.
\end{equation}
We treat the interaction terms $\mathcal{H}_{\mbox{\scriptsize{os}}}$ and $\mathcal{H}_{\mbox{\scriptsize{nn}}}$ in the mean-field approximation. We consider the possibility that Cooper pairs have nonzero center-of-mass momenta, and therefore use an FFLO-type ansatz $U\langle \hat{a}_{j\downarrow}^{\vphantom{\dagger}} \hat{a}_{j\uparrow}^{\vphantom{\dagger}} \rangle = \Delta_0 e^{2i\mathbf{q}\cdot\mathbf{x}_j}$ \cite{notebyPaivi} for the on-site order parameter. Here $\mathbf{x}_j$ is the position vector of lattice site $j$, amplitude $\Delta_0\geq0$ and $2\mathbf{q}$ is the Cooper pair center-of-mass momentum.

On a triangular lattice, there are three different NN bonds. We take the three different NN bonds to be along directions $\mathbf{a}_2$, $\mathbf{a}_1$ and $\mathbf{a}_1-\mathbf{a}_2$ specified in Figs.\ \ref{fig1a} and \ref{fig1b}. We consider a simple situation in which the long-range order parameter has the same norm $\Delta_1$ along all bonds, but different phases are allowed for the different bonds \cite{Shastry}. In equation form, the ansatz reads $V \langle \hat{h}_{mn} \rangle = \Delta_1 e^{i\theta_{mn}} e^{i\mathbf{q}\cdot(\mathbf{x}_m+\mathbf{x}_n)}$, where $\Delta_1\in\mathbb{R}$ and $\theta_{mn}$ is the phase that depends on the direction of the bond between sites $m$ and $n$. We denote the phases corresponding to bonds $\mathbf{a}_2$, $\mathbf{a}_1$ and $\mathbf{a}_1-\mathbf{a}_2$ by $\theta$, $\phi$ and $\varphi$, respectively.

We define the Fourier transformation as $\tilde{f}_{\mathbf{k}\sigma}=M^{-1/2} \sum_j e^{-i\mathbf{k} \cdot \mathbf{x}_j} \hat{f}_{j\sigma}$, where $f\in\lbrace a,b\rbrace$, $\sigma\in\lbrace \uparrow,\downarrow\rbrace$ and $M$ is the number of sites in either of the triangular sublattices A and B. With the help of the Fourier transformation and periodic boundary conditions in $\mathbf{x}$-space, the mean-field Hamiltonian can be written in momentum space as
\begin{eqnarray}
\mathcal{H}_{\mbox{\scriptsize{MF}}} &=& \sum_{\mathbf{k}} \frac{\Delta_0^2}{U} +  \frac{3\Delta_1^2}{V} + \sum_{\alpha=1}^3 \xi^{(\alpha)}_{\mathbf{k}} \hat{c}_{\alpha,\mathbf{k}}^{\dagger} \hat{c}_{\alpha,\mathbf{k}}^{\vphantom{\dagger}} \nonumber \\ 
&&{} + \sum_{\mathbf{k}} \sum_{\beta=1}^2 (g_{\mathbf{k}} + \mathcal{G}_{\mathbf{k}-\mathbf{q}}) \hat{c}_{3,2\mathbf{q}-\mathbf{k}} \hat{c}_{\beta,\mathbf{k}} + \mbox{H.c.}
\end{eqnarray}
in the basis of noninteracting bands. (Triangular lattice Hamiltonian is similar, see Supplemental Material \cite{Omasupp}.) The noninteracting dispersions are explicitly written as $\xi^{(1,2)}_{\mathbf{k}}=\pm |h_{\uparrow}(\mathbf{k})|-\mu_{\uparrow}$, where $h_{\uparrow}(\mathbf{k})=-t_{\uparrow}\lbrack e^{ik_x/\sqrt{3}} + 2e^{-ik_x/(2\sqrt{3})}\cos(k_y/2)\rbrack$ and
$\xi^{(3)}_{\mathbf{k}}=-t_{\downarrow}(2\lbrack \cos k_y +\cos(\lbrack k_y+\sqrt{3}k_x\rbrack/2) + \cos(\lbrack k_y-\sqrt{3}k_x\rbrack/2) \rbrack +3) -\mu_{\downarrow}$. The interband coupling due to the on-site interaction is $g_{\mathbf{k}}=-\Delta_0/\sqrt{2}$. Similarly, the interband coupling due to the NN interaction is $\mathcal{G}_{\mathbf{k}-\mathbf{q}} = -\Delta_1\sum_{\boldsymbol{\delta}} e^{-i\Theta_{\boldsymbol{\delta}}}\cos(\lbrack \mathbf{k}-\mathbf{q}\rbrack\cdot \boldsymbol{\delta})$, where $\sum_{\boldsymbol{\delta}}$ goes over the nearest-neighbors $\mathbf{a}_2$, $\mathbf{a}_1$ and $\mathbf{a}_1-\mathbf{a}_2$, and $\Theta_{\boldsymbol{\delta}}$ is the phase corresponding to $\boldsymbol{\delta}$. When interaction strengths and tunneling amplitudes are fixed, the parameters that govern pairing in the system are the chemical potentials $\mu_{\uparrow}$ and $\mu_{\downarrow}$. The quasiparticle energies $E_{\alpha}(\mathbf{k})$ can be calculated from $\mathcal{H}_{\mbox{\scriptsize{MF}}}$, and the absolute minimum of the grand potential $\Omega(\Delta_0,\Delta_1,\mathbf{q}) = \sum_{\mathbf{k}} \frac{\Delta_0^2}{U} + \frac{3\Delta_1^2}{V} + \xi_3(-\mathbf{k}) -\frac{1}{\beta} \sum_{\alpha=1}^3 \ln (1+e^{-\beta E_{\alpha}(\mathbf{k})})$ determines the values of $\Delta_0$, $\Delta_1$ and $\mathbf{q}$ \cite{Topol}.

A particularly promising way to experimentally realize this model would be to employ the widely used rubidium-potassium mixture composed of fermionic $^{40}$K prepared in the $|F=9/2,m_F=-7/2\rangle$ and $|F=9/2,m_F=-9/2\rangle$ Zeeman components of the $F=9/2$ ground-state hyperfine level and bosonic $^{87}$Rb atoms in the $|F=1,m_F=1\rangle$ ground state. The on-site and NN interactions could be tuned independently \cite{Lim3}, and various experimental methods are available to study the nature of the pairing \cite{TormaBOOK}.

In units of $-(e^2/h)$, the Hall conductance of a filled band is an integer called the Chern number \cite{Bernevig}. If we assume that the pseudo-spin indices $\uparrow$ and $\downarrow$ are associated with internal angular momenta, as opposed to some other internal states unaffected by time reversal, the Hamiltonian $\mathcal{H}$ is not symmetric under time reversal due to the mixed geometry. Despite that, it is easy to show that $\mathcal{H}_{\mbox{\scriptsize{MF}}}$ cannot give rise to phases with a nonzero Chern number if $\theta=\phi=\varphi=0$ and tunneling amplitudes $t_{\uparrow}$ and $t_{\downarrow}$ are real-valued (see Supplemental Material \cite{Omasupp}). In order to study TRS breaking due to the NN interaction, we hereafter say that the pseudo-spin indices $\uparrow$ and $\downarrow$ are not associated with internal angular momenta but by some other internal states unaffected by time reversal. Subsequently, $\mathcal{H}_{\mbox{\scriptsize{MF}}}$ can break TRS only if $\begin{pmatrix} \theta & \phi & \varphi \end{pmatrix} \neq \begin{pmatrix} 0 & 0 & 0 \end{pmatrix}$.

\begin{figure}[h!]
\centering
\subfigure[\ Zero temperature phase diagram as a function of chemical potentials $\mu_{\uparrow}$ and $\mu_{\downarrow}$. The first two main areas are the normal phase and the FFLO phase, while the rest of the phase diagram is covered by various non-FFLO superfluid phases. \label{fig2a}]{
\includegraphics[scale=0.38]{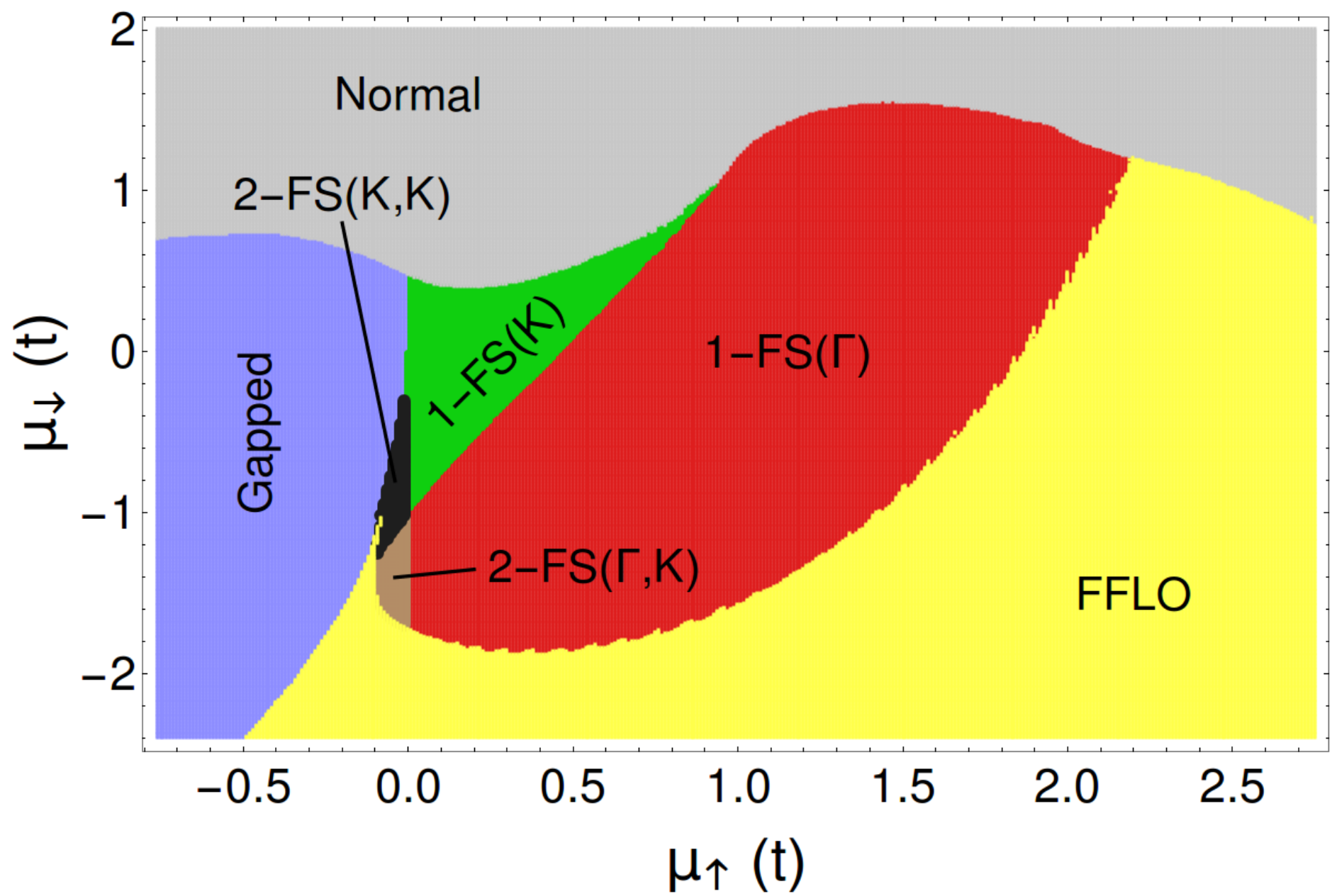}}
\subfigure[\ Density plot of the relative weight of the NN bond $P=|\Delta_1|/(|\Delta_0|+|\Delta_1|)$ as a function of chemical potentials $\mu_{\uparrow}$ and $\mu_{\downarrow}$. \label{fig2b}]{
\includegraphics[scale=0.58]{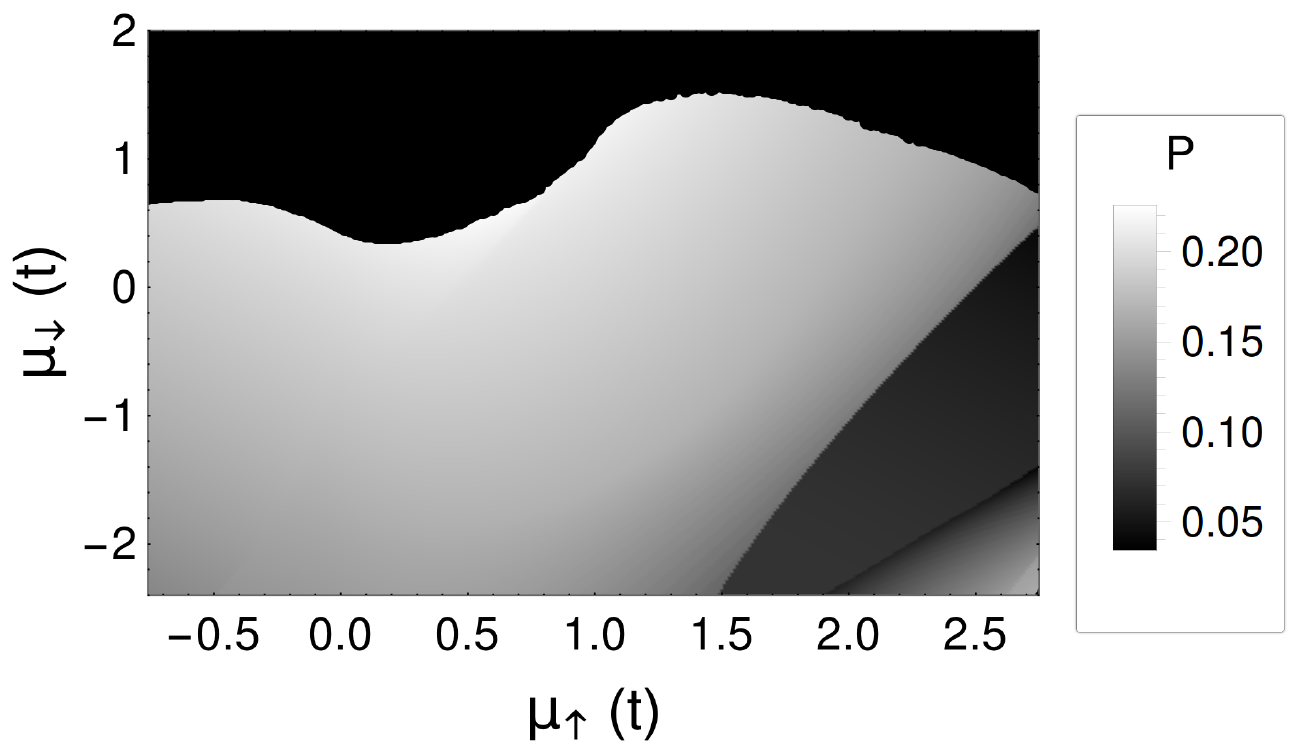}}
\caption{\label{fig2}(Color online.) Phases and pairing when the on-site interaction strength $U=5$ and the NN interaction strength $V=3$.}
\end{figure}

Figure \ref{fig2a} shows the obtained zero temperature phase diagram as a function of $\mu_{\uparrow}$ and $\mu_{\downarrow}$ for $U=5$ and $V=3$. We used the values $\theta=\phi=\varphi=0$ because we have numerically verified that this choice yields the lowest grand potential everywhere except in a small region in the lower right corner of the phase diagram. In other words, the system exhibits phase winding in a small region within the FFLO phase. Moreover, Fig.\ \ref{fig2b} shows that there is significant amount of pairing between nearest-neighbors when $U=5$ and $V=3$. This is very different from the mixed geometry study Ref.\ \cite{Topol} in which long-range interactions were not considered. Moreover, we find a large area of FFLO, which was not included in the ansatz of Ref.\ \cite{Topol}. 

We find that the phase diagram \ref{fig2a} is divided into three main areas. The first two areas are the normal phase and the FFLO superfluid phase, and the third area comprises the rest of the diagram covered by various non-FFLO superfluid phases. The normal phase is simply indicated by vanishing order parameters, i.e.\ $\Delta_0=\Delta_1=0$. On the other hand, FFLO phase is characterized by $\mathbf{q}\neq0$ and at least one of the order parameters $\Delta_0$ and $\Delta_1$ being nonzero. The FFLO phase is an unconventional superfluid phase where Cooper pairs have nonzero center-of-mass momenta. Finally, non-FFLO superfluid phase has $\mathbf{q}=0$ with at least one of the order parameters $\Delta_0$ and $\Delta_1$ being nonzero. The non-FFLO superfluid phase can be further divided into gapless and gapped phases, and the gapless phase can be characterized by the topological arrangement of the one or two Fermi surfaces ($\Gamma$-centered or $K$-centered). The notation 1-FS(X) means 1 Fermi surface centered at high symmetry point X and notation 2-FS(X,Y) means 2 Fermi surfaces centered at high symmetry points X and Y \cite{Topol}. However, since $\theta=\phi=\varphi=0$, the phases necessarily have vanishing Chern numbers. 

\begin{figure}[h!]
\centering
\subfigure{\includegraphics[scale=0.50]{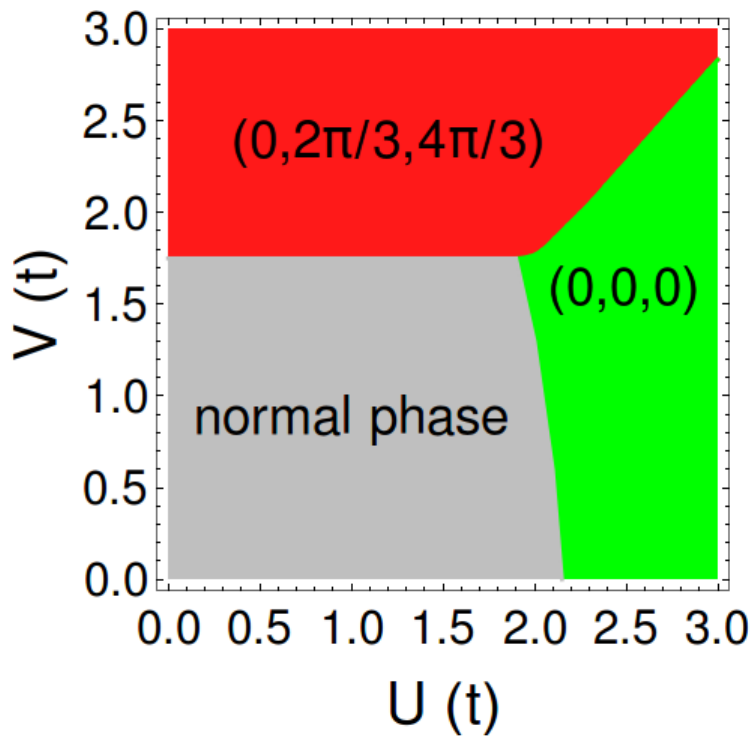}\label{fig3a}}
\subfigure{\includegraphics[scale=0.48]{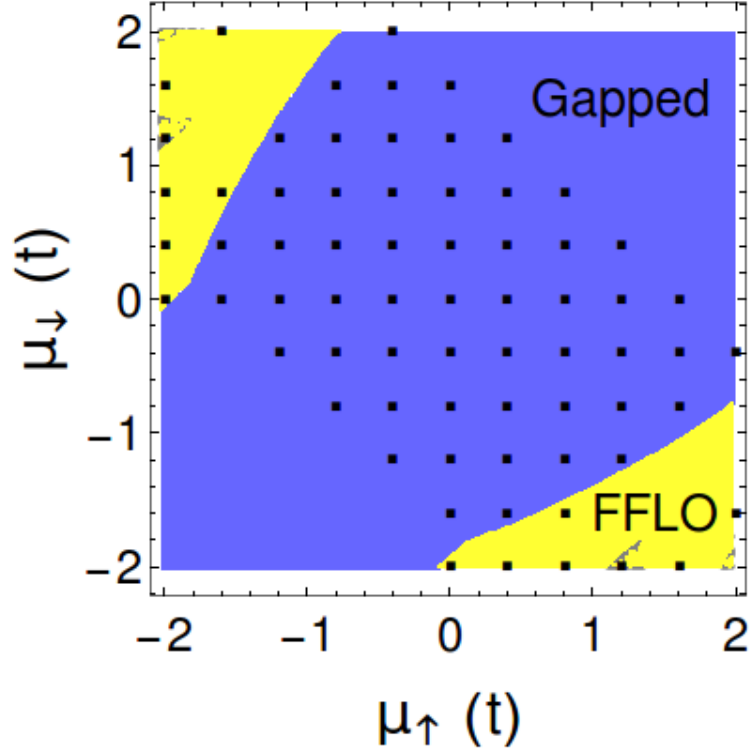}\label{fig3b}}
\caption{(Color online.) a) Honeycomb-triangular lattice phase angles $\begin{pmatrix} \theta & \phi & \varphi \end{pmatrix}$ at the point $\begin{pmatrix} \mu_{\uparrow} & \mu_{\downarrow} \end{pmatrix} = \begin{pmatrix} -1.5 & -2.5 \end{pmatrix}$ as a function of on-site and NN interaction strengths $U$ and $V$. b) Triangular lattice phase diagram for $U=V=5$ with $\begin{pmatrix} \theta & \phi & \varphi \end{pmatrix} = \begin{pmatrix} 4\pi/3 & 2\pi/3 & 0 \end{pmatrix}$. Black squares indicate the area where the grand potential is minimized by $\begin{pmatrix} \theta & \phi & \varphi \end{pmatrix} = \begin{pmatrix} 4\pi/3 & 2\pi/3 & 0 \end{pmatrix}$.}
\end{figure}

Now, it is of interest to ask whether the system breaks TRS for some values of $U$, $V$, $\mu_{\uparrow}$ and $\mu_{\downarrow}$. To that end, Fig.\ \ref{fig3a} shows the phase angles $\theta$, $\phi$ and $\varphi$ as a function of $U$ and $V$ at the point $\begin{pmatrix} \mu_{\uparrow} & \mu_{\downarrow} \end{pmatrix} = \begin{pmatrix} -1.5 & -2.5 \end{pmatrix}$. Temperature was set to zero. At lower values of $U$ the system is in normal phase if $V$ is small and in superfluid phase with $\begin{pmatrix} \theta & \phi & \varphi \end{pmatrix} = \begin{pmatrix} 0 & 2\pi/3 & 4\pi/3 \end{pmatrix}$ if $V$ is large. At higher values of $U$ the system is in superfluid phase with $\begin{pmatrix} \theta & \phi & \varphi \end{pmatrix} = \begin{pmatrix} 0 & 0 & 0 \end{pmatrix}$ if $V$ is small, and in superfluid phase with $\begin{pmatrix} \theta & \phi & \varphi \end{pmatrix} = \begin{pmatrix} 0 & 2\pi/3 & 4\pi/3 \end{pmatrix}$ for large values of $V$. {\it{Thus the system spontaneously breaks TRS when $V$ becomes large enough.}} We also note that the threshold for TRS breaking becomes higher when $U$ is raised. TRS breaking also happens in the triangular lattice \cite{Shastry}, but the phase diagram shown in Fig.\ \ref{fig3b} is exceedingly simple compared to the rich phase diagram of Fig.\ \ref{fig2a}.

\begin{figure}[h!]
\centering
\includegraphics[scale=0.7]{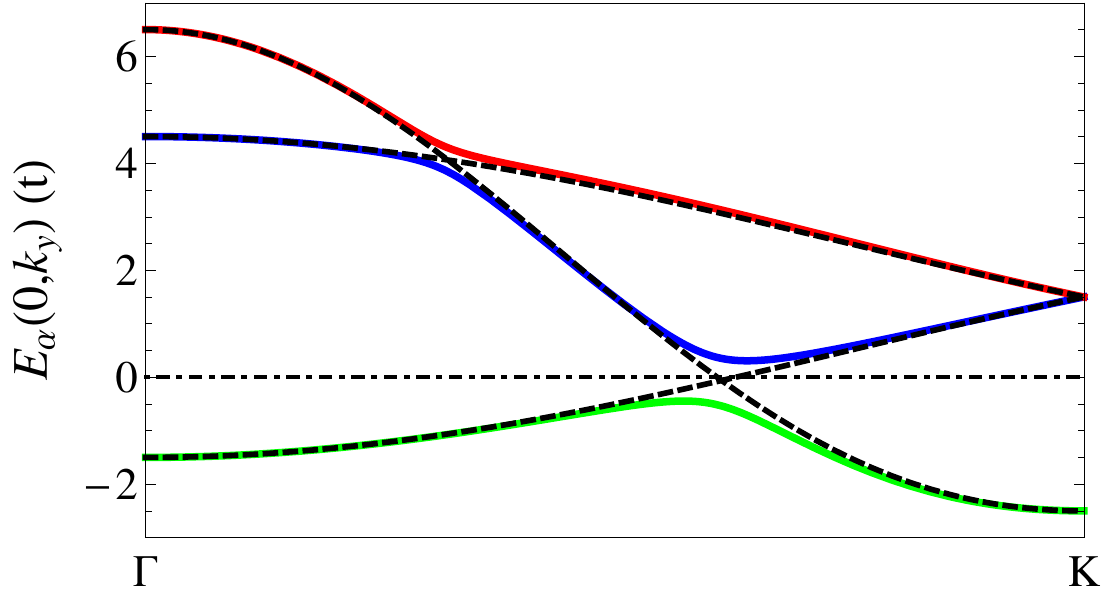}
\captionsetup{singlelinecheck=off}
\caption[.]{(Color online.) Quasiparticle energy bands $E_{\alpha}(k_x,k_y)$ on the line $\Gamma$--$K$ for interacting (solid) and noninteracting (dashed) systems when $\begin{pmatrix} \mu_{\uparrow} & \mu_{\downarrow} \end{pmatrix} = \begin{pmatrix} -1.5 & -2.5 \end{pmatrix}$ and $U=0$ and $V=3$. Dash-dotted line indicates the Fermi energy $E_{\mbox{\scriptsize{F}}}=0$.}
\label{fig4}
\end{figure}

Figure \ref{fig4} shows the quasiparticle energy bands $E_1(\mathbf{k})$, $E_2(\mathbf{k})$ and $E_3(\mathbf{k})$ along the line $\Gamma$--$K$ for the point $\begin{pmatrix} \mu_{\uparrow} & \mu_{\downarrow} \end{pmatrix} = \begin{pmatrix} -1.5 & -2.5 \end{pmatrix}$ when $U=0$ and $V=3$. The system is in a gapped phase because none of the energy bands cross the Fermi level located at $E_{\mbox{\scriptsize{F}}}=0$. In addition, we note that the two higher bands are degenerate at the Dirac points $K$ because the coupling function $\mathcal{G}_{\mathbf{k}}$ vanishes at the Dirac points. 

\noindent We have calculated the Chern numbers by using the method from Ref.\ \cite{Fukui}. In that method, one obtains the Chern number by summing a gauge independent field strength $\mathcal{F}_{12}(\mathbf{k}_l)$ over a set of discrete points $\mathbf{k}_l$ covering the entire Brillouin zone. Due to the periodicity of the $\mathbf{k}$-space Hamiltonian, the Brillouin zone can be regarded as a two-dimensional torus. Remarkably, the field strengths $\mathcal{F}_{12}(\mathbf{k}_l)$ can also be directly measured by using time-of-flight imaging \cite{Deng}. We found that the Chern number for the lowest band is $c_3=2$. However, the two higher bands do not satisfy the gap opening condition $|E_1-E_2|\neq0$ at the Dirac points $K$. Therefore we did not calculate the Chern numbers for those bands individually, but for the multiplet $\psi$ comprising the two bands. The multiplet Chern number $c_{\psi}=-2$. Although we have calculated the Chern numbers using periodic boundary conditions, the nonzero Chern numbers still suggest that a finite system with edges would have propagating edge modes \cite{Hatsugai1,Hatsugai2}. The main challenge in detecting such edge modes has been the separation of the small edge-state signal from the bulk background, but Ref.\ \cite{Goldman} provides a simple and robust way to measure the edge modes. Moreover, when the Fermi energy lies in a gap, the Hall conductance is given by $\sigma_{xy}=-(e^2/h)\sum_n c_n$, where $c_n$ denotes the Chern number of the $n$th Bloch band and the sum over $n$ is restricted to the bands below the Fermi energy \cite{Fukui,Thouless,Kohmoto}. The lowest energy band in Fig.\ \ref{fig4} is fully below the Fermi energy $E_{\mbox{\scriptsize{F}}}=0$, whereas the two higher bands are completely above the Fermi energy. Consequently, the Hall conductance is $-c_3=-2$ in units of $e^2/h$.

It is remarkable that {\it simultaneous} occurrence of phase winding and FFLO is possible both in honeycomb-triangular and triangular lattices. In a honeycomb-triangular lattice time-reversal and translational symmetries are simultaneously broken e.g.\ at $\begin{pmatrix} \mu_{\uparrow} & \mu_{\downarrow} \end{pmatrix} = \begin{pmatrix} 2 & -2 \end{pmatrix}$ when $U=0$ and $V=4$, whereas Fig.\ \ref{fig3b} shows the areas where this happens in a triangular lattice for $U=V=5$. Although it is known that TRS can be broken in a triangular lattice due to NN interactions \cite{Shastry}, we have shown here that {\it simultaneous} breaking of time-reversal and translational symmetries in the superfluid order parameter of a two-component fermion system may happen both in honeycomb-triangular and triangular lattices. 

In summary, the extended Fermi-Hubbard model we have considered in a mixed honeycomb-triangular lattice exhibits a rich phase diagram with gapped and gapless paired phases, as well as spontaneous TRS breaking at NN interaction strengths $V$ higher or equal to the on-site interaction $U$. The TRS breaking gives rise to topologically nontrivial phases and nonzero Hall conductivity. The connection of our lattice model to various graphene systems \cite{Nandkishore, Jiang, Uchoa} may inspire a search for ways to design mixed geometries on such nanomaterials. Remarkably, we found that TRS breaking happens also in the FFLO state: we thus predict a novel type of superfluid with simultaneous TRS and translational symmetry breaking. This new phase of matter could be realized in the mixed honeycomb-triangular or in the triangular geometry which are both realizable in ultracold gases, the latter being simpler since it does not require spin-dependent confinement.

\begin{acknowledgments}
We thank D.-H.\ Kim for useful discussions. This work was supported by the Academy of Finland through its Centres of Excellence Programme (2012-2017) and under Project Nos.\ 263347, 251748 and 272490, and by the European Research Council (ERC-2013-AdG-340748-CODE). We acknowledge the computational resources provided by Aalto University Science-IT project. One of us (R.S.) acknowledges a grant from the University of Oulu.
\end{acknowledgments}

\bibliography{bibliography}

\begin{thebibliography}{61}
\bibitem [{not()}]{notebyPaivi}%
  \BibitemOpen
  \href@noop {} {\bibinfo  {journal} {When such an FF ansatz minimizes the
  energy, the actual ground state may be an LO solution of cosine form where
  the translational symmetry is broken in the amplitude and not only phase of
  the order parameter. The LO solutions in known cases have lower energies than
  the FF, thus FF ground states can be taken as indicators of more general
  FFLO-type states}\ }\BibitemShut {NoStop}%
\end{thebibliography}
\end{document}